\documentclass{jetpl_arXiv09074887v6}
\twocolumn
\title{Towards a solution of the cosmological constant problem}

\rtitle{Towards a solution of the cosmological constant problem}

\sodtitle{Towards a solution of the cosmological constant problem}

\author{F.R. Klinkhamer $^{\#}$\/$^{1)}$
and
G.E. Volovik $^{*+}$\/\thanks{\:e-mail: frans.klinkhamer@kit.edu, volovik@boojum.hut.fi}}

\rauthor{F.R. Klinkhamer, G.E.Volovik}

\sodauthor{Klinkhamer, Volovik}

\address{$^{\#}$ Institute for Theoretical Physics, University of
Karlsruhe, Karlsruhe Institute of Technology, 76128 Karlsruhe, Germany
\\
$^{*}$ Low Temperature Laboratory, Aalto University, FI-00076 AALTO, Finland
\\
$^+$ Landau Institute for Theoretical Physics RAS, Kosygina 2,
119334 Moscow, Russia
  }

\dates{January 25, 2010}{*}

\PACS{04.20.Cv, 98.80.Es, 95.36.+x}

\abstract{The standard model of elementary particle physics
and the theory of general relativity can be extended by
the introduction of a vacuum variable which is responsible for
the near vanishing of the present cosmological constant (vacuum energy density).
The explicit realization of this vacuum variable can be via
a three-form gauge field, an aether-type velocity field, or any other
field appropriate for the description of the equilibrium state
corresponding to the Lorentz-invariant quantum vacuum.
The extended theory has, without fine-tuning,
a Minkowski-type solution of the field equations
with spacetime-independent fields and provides, therefore,
a possible solution of the main cosmological constant problem.}

\begin{document}

\maketitle

\newcommand  {\version}{v5.99}

% Macros for text:
\newcommand{\beq}{\begin{equation}}
\newcommand{\eeq}{\end{equation}}
\newcommand{\beqa}{\begin{eqnarray}}
\newcommand{\eeqa}{\end{eqnarray}}
\newcommand{\bsubeqs}{\begin{subequations}}
\newcommand{\esubeqs}{\end{subequations}}
\newcommand{\dd}{\mathrm{d}}                    % differential d
\newcommand{\half}{{\textstyle \frac{1}{2}}}    % small fraction 1/2

\section{Introduction}\label{sec:intro}

The main cosmological constant problem is to understand why, naturally, the
quantum-mechanical
zero-point energy of the vacuum does not produce a large cosmological constant
or, in other words, to discover the way the zero-point energy is canceled
without fine-tuning the theory.
Restricting to established physics, this problem was formulated by Weinberg
in the following pragmatic way~\cite{Weinberg1988,Weinberg1996}:
how to find an extension of the standard model
of elementary particle physics and the theory of  general relativity,
for which  there exists, without fine-tuning,  a Minkowski-spacetime
solution with spacetime-independent fields.

An adjustment-type solution of the cosmological constant problem
appears, however, to be impossible with a fundamental scalar field and
Weinberg writes in the last sentence of Sec.~2 in Ref.~\cite{Weinberg1996}
that, to the best of his knowledge, ``no one has found a way out of this impasse.''
In this Letter, we present a way \emph{around} the impasse, which employs a
quantity $q$ that acts as a self-adjusting scalar field but is
non-fundamental~\cite{KlinkhamerVolovik2008a,KlinkhamerVolovik2008b,KlinkhamerVolovik2009b}.

The main goal of the present publication is to describe,
in a more or less consistent way, a particular theoretical framework
for addressing the cosmological constant problem.
Obviously, this builds on previous work of the present authors
and many others (see citations below). But there are also two important
new results, which will be indicated explicitly.

\section{Minkowski equilibrium vacuum}\label{sec:Statics}

Our discussion starts from the theory outlined in
Ref.~\cite{KlinkhamerVolovik2008b}. We introduce a special quantity,
the  vacuum ``charge''  $q$, to describe the statics of the quantum vacuum.
A concrete example of this vacuum variable is given by the four-form field
strength~\cite{DuffNieuwenhuizen1980,Aurilia-etal1980,Hawking1984,HenneauxTeitelboim1984,
Duff1989,DuncanJensen1989,BoussoPolchinski2000,Aurilia-etal2004,Wu2008} expressed in
terms of $q$ as $F_{\alpha\beta\gamma\delta}=q\,\sqrt{-\det g}\,
\epsilon_{\alpha\beta\gamma\delta}$ (see below for further details).
This particular vacuum variable $q$ is associated with an
energy scale $E_\text{UV}$ that is assumed to be much larger than the
electroweak energy scale $E_\text{ew} \sim 10^{3}\,\text{GeV}$
and possibly to be of the order of the
gravitational energy scale $E_\text{Planck} \equiv 1/\sqrt{8\pi G_{N}}
\approx 2.44\times 10^{18}\:\text{GeV}$. Here, and in the following,
natural units are used with $\hbar=c=1$.

Specifically, the effective action of our theory is given by
\bsubeqs\label{eq:EinsteinF-all}
\beqa
\hspace*{-5mm}
S^\text{eff}[A, g,\psi]
&=&
-  \int_{\mathbb{R}^4}\,d^4x\, \sqrt{-\det g}\, \Big(K(q)\,R[g]
\nonumber\\
\hspace*{-5mm}&&
+\epsilon(q)  +\mathcal{L}_\text{SM}^\text{eff}[\psi,g]\Big)\,,
\label{eq:actionF}\\[2mm]
\hspace*{-5mm}
F_{\alpha\beta\gamma\delta}  &\equiv&
q\,\epsilon_{\alpha\beta\gamma\delta}\, \sqrt{-\det g}
= \nabla_{[\alpha}A_{\beta\gamma\delta]}\,,
\label{eq:Fdefinition}\\[2mm]
q^2 &=&
- \frac{1}{24}\,F_{\alpha\beta\gamma\delta}\,F^{\alpha\beta\gamma\delta}\,,
\label{eq:q2definition}
\eeqa
\esubeqs
where $R$ denotes the Ricci curvature scalar,
$\epsilon_{\alpha\beta\gamma\delta}$ the Levi--Civita tensor density,
$\nabla_\alpha$ the covariant derivative,
and the square bracket around spacetime indices complete anti-symmetrization.
Throughout, we use the same conventions as in Ref.~\cite{Weinberg1988},
in particular, those for the Riemann curvature tensor
and the metric signature $(-+++)$.

The vacuum energy density $\epsilon$ in \eqref{eq:actionF}
depends on the vacuum variable $q=q[A,g]$ and
the same is assumed to hold for the gravitational coupling parameter $K$.
The single field $\psi$ combines all the fields of the standard model
(spinor, gauge, Higgs, and ghost fields~\cite{Veltman1994})
and, for simplicity, the scalar Lagrange density
$\mathcal{L}_\text{SM}^\text{eff}$ in \eqref{eq:actionF}
is taken to be without direct $q$ dependence.
The original standard model fields collected in $\psi(x)$ are quantum fields
with vanishing vacuum expectation values in Minkowski spacetime
(this holds, in particular, for the physical Higgs field $H(x)$~\cite{Veltman1994}).
The effective action takes $\psi(x)$ to be a classical field,
but has additional terms to reflect the quantum effects~\cite{BirrellDavies1982}.
The metric field $g_{\alpha\beta}(x)$
and the three-form gauge field $A_{\beta\gamma\delta}(x)$
[or other $q$--related fields discussed later on]
are, for the moment, considered to be genuine classical fields.

The setup, now, is such that a possible constant term $\Lambda_\text{SM}$ in
$\mathcal{L}_\text{SM}^\text{eff}$ (which includes the zero-point energies
from the standard model fields) has been absorbed in $\epsilon(q)$,
so that, in the end, $\mathcal{L}_\text{SM}^\text{eff}[\psi,g]$ contains only
$\psi$--dependent terms, with the metric $g_{\alpha\beta}$
(or vierbein $e_{\alpha}^{a}$)
entering through the usual covariant derivatives.
In short, the following holds true:
\beq
\mathcal{L}_\text{SM}^\text{eff}[\psi_0,\eta]=0\,,
\label{eq:Leff}
\eeq
where $\psi_0$ denotes the constant values for the standard model
fields over  Minkowski spacetime and $\eta$ stands for the Minkowski metric
$\eta_{\alpha\beta} =\text{diag}(-1,\, 1,\, 1,\, 1)$ in standard coordinates.

The actual spectrum of the vacuum energy density (meaning the
different contributions to $\epsilon$ from different energy scales)
is not important for the cancellation mechanism to be discussed in this Letter.
Still, we assume, for definiteness, that the vacuum energy density $\epsilon(q)$
splits into a constant part and a variable part:
\beq
\epsilon(q)=      \Lambda_\text{bare} + \epsilon_\text{var}(q)
           \equiv \Lambda_\text{SM} + \Lambda_\text{UV} + \epsilon_\text{var}(q)\,,
\label{eq:general-epsilon}
\eeq
with $\partial \epsilon_\text{var}/\partial q \ne 0$,
a constant term $\Lambda_\text{SM}$ of typical size
$|\Lambda_\text{SM}| \sim (E_\text{ew})^4$ removed from
$\mathcal{L}_\text{SM}^\text{eff}$ according to \eqref{eq:Leff}, and
a possible extra contribution $\Lambda_\text{UV}$
of size $|\Lambda_\text{UV}| \sim (E_\text{UV})^4$
from the unknown physics beyond the standard model.
For definiteness, we also assume that $\epsilon_\text{var}(q)$ contains only
even powers of $q$ and recall that $q^2$  is defined
by \eqref{eq:q2definition} in terms of the three-form gauge field $A$
entering the field strength \eqref{eq:Fdefinition}.
Allowing for a general even function $\epsilon(q)$ instead of
the single Maxwell-type term $\half\,q^2$ considered in the previous
literature~\cite{DuffNieuwenhuizen1980,Aurilia-etal1980,Hawking1984,HenneauxTeitelboim1984,Duff1989}
will turn out to be an important ingredient for the cancellation of
$\Lambda_\text{bare}$ values of arbitrary sign.

The generalized Maxwell and Einstein equations from action \eqref{eq:actionF}
have been derived in Ref.~\cite{KlinkhamerVolovik2008b}.
The generalized Maxwell equation reads
\begin{equation}
\nabla_\alpha \left(\sqrt{-\det g} \,\;\frac{F^{\alpha\beta\gamma\delta}}{q}
           \left(\frac{\partial\epsilon(q)}{\partial q}+R\;\frac{\partial K(q)}
            {\partial q} \right) \right)=0
\label{eq:genMaxwell}
\end{equation}
and reproduces the known equation~\cite{DuffNieuwenhuizen1980,Aurilia-etal1980}
for the special case $\epsilon(q)=\half\,q^2$ and $\partial K/\partial q=0$.
The first integral of \eqref{eq:genMaxwell} with integration constant $\mu$
and the final version of the generalized Einstein equation
then give the following generic equations~\cite{KlinkhamerVolovik2008b}:
\bsubeqs\label{eq:genMaxwellSolutionEinsteinEquation}
\begin{eqnarray}
\hspace*{-5mm}
\frac{\partial\epsilon(q)}{\partial q}
+ R\,\frac{\partial K(q)}{\partial q}
&=&\mu\,,
\label{eq:genMaxwellSolution}\\[2mm]
\hspace*{-5mm}
 2K\,\big( R^{\alpha\beta}-g^{\alpha\beta}\,R/2 \big)
&=&
-2\,\big(\nabla^\alpha\nabla^\beta - g^{\alpha\beta}\, \square\big)\, K(q)
\nonumber\\[1mm]
\hspace*{-5mm}
&&
+\big(\epsilon(q)-  \mu \, q \big)\, g^{\alpha\beta} -T^{\alpha\beta}_\text{SM}\,,
\label{eq:genEinsteinEquation}
\end{eqnarray}
\esubeqs
where $T^{\alpha\beta}_\text{SM}$ is the
energy-momentum tensor corresponding to the effective Lagrangian
appearing in \eqref{eq:actionF} and \eqref{eq:Leff}. From general coordinate
invariance, the energy-momentum tensor is known to have a vanishing covariant
divergence, $\nabla_\alpha\,T^{\alpha\beta}_\text{SM}=0$.

For the special case $K(q)=K_0=\text{const}$, \eqref{eq:genEinsteinEquation}
reduces to the standard Einstein equation of general relativity.
For the general case $dK/dq\ne 0$, the action \eqref{eq:actionF}
and the resulting field equation \eqref{eq:genEinsteinEquation}
correspond to those of Brans--Dicke theory \cite{BransDicke1961},
but without kinetic term for the scalar degree of freedom
($\omega_\text{BD}=0$). See also the related work on
inflation theory~\cite{Starobinsky1980},
dark-energy models~\cite{HuSawicki2007,ApplebyBattye2007,Starobinsky2007,Brax-etal2008},
and the connection
to $q$--theory~\cite{KlinkhamerVolovik2008jetpl,KlinkhamerVolovik2009a,Klinkhamer2009}.

The crucial difference between our theory and conventional $f(R)$ modified-gravity
theories~\cite{Starobinsky1980,HuSawicki2007,ApplebyBattye2007,Starobinsky2007,Brax-etal2008}
lies in the appearance, for us, of the integration constant $\mu$
after integration over the three-form gauge field $A$,
i.e., after solving the generalized Maxwell equation \eqref{eq:genMaxwell}.
As a result, the vacuum energy density entering the
generalized Einstein equation \eqref{eq:genEinsteinEquation}
is not the original vacuum energy density $\epsilon(q)$ from the
action \eqref{eq:actionF} but the combination
\beq
 \rho_{V}(q) \equiv \epsilon(q)-  \mu \, q \,.
\label{eq:rho_V}
\eeq
This gravitating vacuum energy density
becomes a genuine cosmological constant
$\overline{\Lambda}\equiv \Lambda(\overline{q}) = \rho_{V}(\overline{q})$
for a spacetime-independent vacuum variable $\overline{q}$.

The field equations (\ref{eq:genMaxwellSolutionEinsteinEquation}ab)
can now be seen to have a Minkowski-type solution with
spacetime-independent fields. For standard global spacetime coordinates,
the fields of this constant solution are given by
\bsubeqs\label{eq:Minkowski-type-solution}
\beqa
g_{\alpha\beta}(x) &=& \eta_{\alpha\beta}\,,\\
F_{\alpha\beta\gamma\delta}(x)&=&q_0\,\epsilon_{\alpha\beta\gamma\delta}\,,\\
\psi(x)       &=& \psi_0\,,
\eeqa
\esubeqs
with spacetime-independent parameters $\mu_0$ and $q_0$ determined
by the following two conditions:
\bsubeqs\label{eq:equil-eqs}
\beqa
\Bigg[\, \frac{\dd \epsilon(q)}{\dd q} - \mu\, \Bigg]_{\mu=\mu_0\,,\,q=q_0} &=&0\,,
\label{eq:equil-eqs-mu}
\\[2mm]
\Big[\, \epsilon(q)    -  \mu\, q\,\Big]_{\mu=\mu_0\,,\,q=q_0} &=&0\,.
\label{eq:equil-eqs-GDcondition}
\eeqa \esubeqs
Conditions \eqref{eq:equil-eqs-mu} and \eqref{eq:equil-eqs-GDcondition}
follow from \eqref{eq:genMaxwellSolution} and \eqref{eq:genEinsteinEquation},
respectively, for $R=R^{\alpha\beta}= T^{\alpha\beta}_\text{SM}=0$
and spacetime-independent $q_0$.

The two conditions \eqref{eq:equil-eqs-mu}--\eqref{eq:equil-eqs-GDcondition}
can be combined into a  \emph{single} equilibrium condition for $q_0$:
\beq
\Lambda_0 \equiv
\Bigg[\, \epsilon(q)  -  q\,\frac{\dd \epsilon(q)}{\dd q}\, \Bigg]_{\,q=q_0} = 0\,,
\label{eq:equil-eqs-q0}
\eeq
with the \emph{derived} quantity~\cite{endnote:chemical-potential}
\beq
\mu_0 = \Bigg[\, \frac{\dd \epsilon(q)}{\dd q}\,    \Bigg]_{\,q=q_0}\,.
\label{eq:equil-eqs-mu0}
\eeq
The spacetime independence of $q_0$ implies that of
$\mu_0$ in \eqref{eq:equil-eqs-mu0} and, with \eqref{eq:genMaxwellSolution},
guarantees that the
generalized Maxwell equation \eqref{eq:genMaxwell} is automatically
solved by the Minkowski-type solution \eqref{eq:Minkowski-type-solution};
see below for a general discussion of this important point.
In order for the Minkowski vacuum to be stable, there is the further condition:
\begin{equation}
\big(\chi_0\big)^{-1} \equiv
\Bigg[\, q^2\;\frac{\dd^2  \epsilon(q)}{\dd  q^2}\,\Bigg]_{q=q_0} > 0\,,
\label{eq:chi_0}
\end{equation}
where $\chi$ corresponds to the isothermal vacuum
compressibility~\cite{KlinkhamerVolovik2008a}.
In the equilibrium vacuum relevant to our Universe,
the gravitational constant $K(q_0)$
of the action \eqref{eq:actionF} can be identified
with $K_0\equiv 1/\big(16\pi\,G_{N}\big)$ in terms of
Newton's constant $G_{N}$.

\begin{figure}[t]   %%   %%
\vspace*{6mm}
\begin{center}
\includegraphics[width=0.35\textwidth]{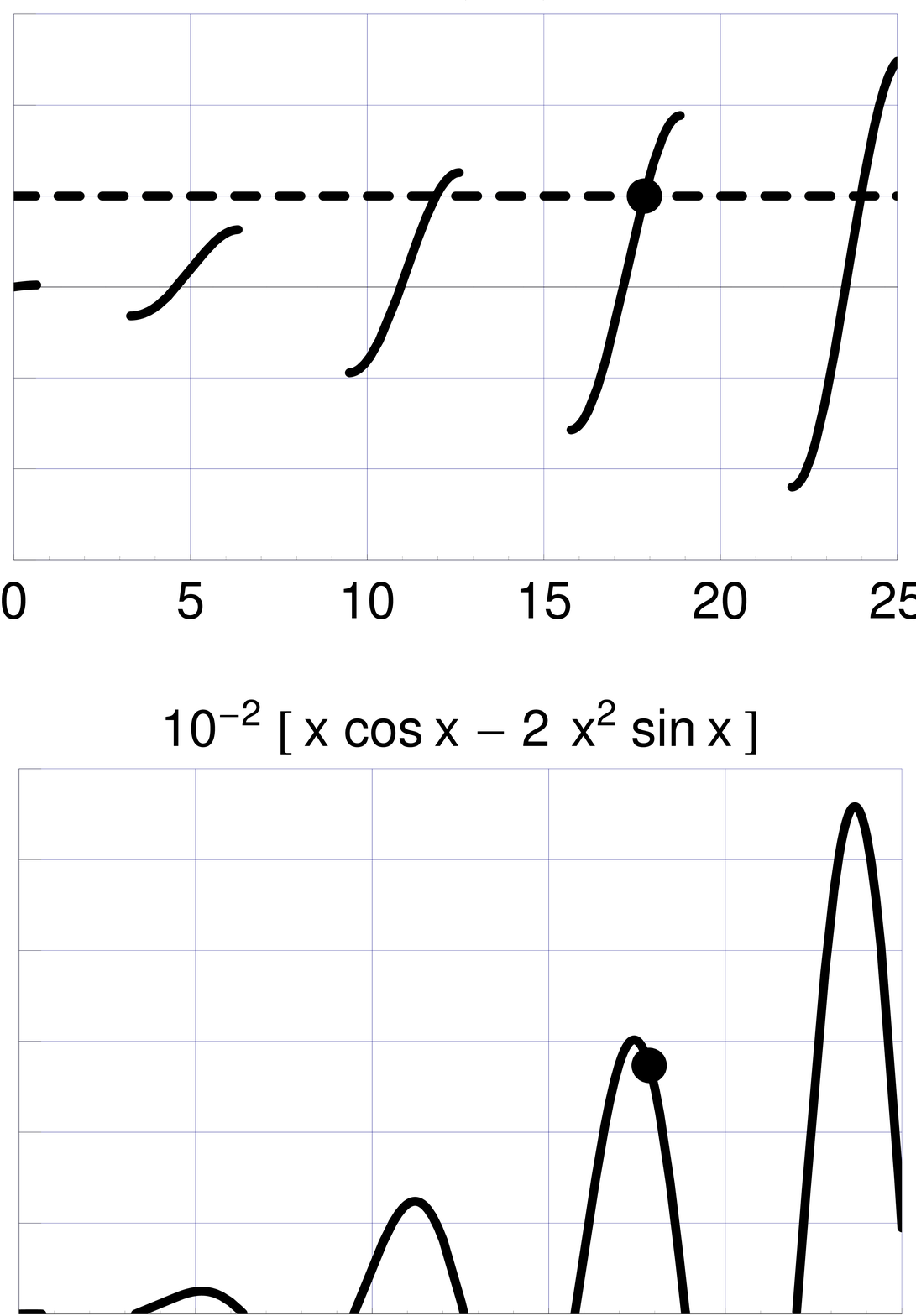}
\end{center}
\vspace*{4mm}
\caption{\textbf{Fig.~1~}
Minkowski equilibrium vacua for a particular choice of
the vacuum energy density function as given by \eqref{eq:epsilon-example}.
The curves of the top panel show the left-hand side
of \eqref{eq:q0-eq} for those
values of $x \equiv \widehat{q}^{\;2}$ that obey the stability condition
\eqref{eq:q0-cond}. The curves of the bottom panel show the corresponding
positive segments of the inverse of the dimensionless vacuum
compressibility $\widehat{\chi}$ defined by the left-hand side of
\eqref{eq:q0-cond}, the general dimensional quantity being defined by
\eqref{eq:chi_0}. Minkowski-type vacua \eqref{eq:Minkowski-type-solution}
are obtained at the intersection points of the curves of the top panel with
a horizontal line at the value $\lambda\equiv \Lambda_\text{bare}/(E_\text{UV})^4$
[for example, the dashed line at
$\lambda=10$ gives the value $(\widehat{q}_0)^2 \approx 17.8453$
corresponding to the heavy dot in the top panel]. Each such vacuum is
characterized, in part, by the corresponding value of the inverse vacuum
compressibility from the bottom panel [for example,
$1/\widehat{\chi}_0 \approx 546.974$
shown by the heavy dot for the case chosen in the top panel].
Minkowski vacua with positive compressibility are stable and become
attractors in a dynamical context (cf. Sec.~3 and Fig.~2).
}
\label{fig:GibbsDuhemLHSandChi}
\end{figure}

Equation~\eqref{eq:equil-eqs-q0} corresponds to the
first of the two constant-field equilibrium conditions
given by Weinberg~\cite{Weinberg1988} as Eqs.~(6.2) and (6.3):
$\partial \mathcal{L} / \partial g_{\alpha\beta}=0$
and $\partial \mathcal{L} / \partial \phi =0$, having
restricted the discussion here to the case of a single
fundamental scalar field $\phi$.
These two conditions turn out to be inconsistent,
unless the potential term in $\mathcal{L}(\phi)$ is fine-tuned~\cite{Weinberg1988}.
See also Sec.~2 of Ref.~\cite{Weinberg1996} for further discussion on
the impossibility of finding a natural Minkowski-type
solution from the adjustment of a fundamental scalar field.

The crucial difference between a fundamental scalar field $\phi$
and our vacuum variable $q$ (a non-fundamental scalar field)
is that the equilibrium condition for $q$ is \emph{relaxed}:
we find, instead of the condition $\partial \mathcal{L} / \partial q =0$,
the conditions $\nabla_\alpha (\partial \mathcal{L} / \partial q)=0$,
which allow for having $\partial \mathcal{L} / \partial q=\mu$
with an \emph{arbitrary} constant $\mu$. As a result, the
equilibrium conditions for $g_{\alpha\beta}$ and $q$ can be consistent
without fine-tuning. The approach based on such a $q$--variable
bypasses the apparent no-go theorem
(as foretold by Ftn.~8 of Ref.~\cite{Weinberg1988}) and formally
solves the cosmological constant problem
(as formulated in Sec.~2 of Ref~\cite{Weinberg1996}):
the original action is not fine-tuned and need not vanish
at the stationary point, but there still exists a Minkowski-type solution
of the field equations.
This realization that the $q$--field allows for a relaxation of
the equilibrium condition is the first of the two most important new results
of the present Letter.

The Minkowski-type solution of theory \eqref{eq:EinsteinF-all}
is given by the fields \eqref{eq:Minkowski-type-solution} with
a constant $q_0$ parameter that solves \eqref{eq:equil-eqs-q0} and
satisfies \eqref{eq:chi_0}. At this moment, it may be instructive to work out a
concrete example. A particular choice for the vacuum energy density
function \eqref{eq:general-epsilon} is given by:
\beq
\epsilon(q)= \Lambda_\text{bare} + (1/2)\,(E_\text{UV})^4\,
\sin\big[\,q^2/(E_\text{UV})^4\,\big]\,,
\label{eq:epsilon-example}
\eeq
which contains higher-order terms in addition to the standard
quadratic term $\half\,q^2$.
Needless to say, many other functions $\epsilon(q)$ can be chosen,
the only requirement being that the equilibrium and stability conditions
can be satisfied~\cite{KlinkhamerVolovik2008a}.
With \eqref{eq:epsilon-example}, the expressions for
the equilibrium condition \eqref{eq:equil-eqs-q0} and
the stability condition \eqref{eq:chi_0} become
\bsubeqs\label{eq:q0-eq-cond}
\beqa
\widehat{q}^{\;2} \,\cos\big(\widehat{q}^{\;2}\big)
-(1/2)\,\sin\big(\widehat{q}^{\;2}\big)                    &=& \lambda \,,
\label{eq:q0-eq}\\[3mm]
\widehat{\chi}^{\;-1} \equiv
\widehat{q}^{\;2}\cos\big(\widehat{q}^{\;2}\big)
-2\, \widehat{q}^{\;4}\sin\big(\widehat{q}^{\;2}\big)    &>& 0\,,
\label{eq:q0-cond}
\eeqa
\esubeqs
where $E_\text{UV}$ has been used to define dimensionless quantities
$\widehat{q} \equiv q/(E_\text{UV})^2$
and $\lambda\equiv \Lambda_\text{bare}/(E_\text{UV})^4$.
A straightforward  graphical analysis (Fig.~\ref{fig:GibbsDuhemLHSandChi})
shows that, for any $\lambda \in \mathbb{R}$,
there are infinitely many values $\widehat{q}_0 \in \mathbb{R}$
which obey both \eqref{eq:q0-eq} and \eqref{eq:q0-cond}. The top panel of
Fig.~\ref{fig:GibbsDuhemLHSandChi} also shows that the $\widehat{q}$ values
on the one segment singled-out by the heavy dot already allow for a complete
cancellation of \emph{any} $\Lambda_\text{bare}$ value
between $-15\,(E_\text{UV})^4$ and $+18\,(E_\text{UV})^4$.

\section{Minkowski attractor}\label{sec:Dynamics}

The cancellation mechanism discussed in the previous section
provides the following general lesson.
The Minkowski-type solution \eqref{eq:Minkowski-type-solution}
appears without fine-tuning of the parameters of the action,
precisely because the vacuum is characterized by a constant derivative
of the vacuum field
rather than by a constant value of the vacuum field itself.
As a result, the parameter $\mu_0$ emerges in \eqref{eq:equil-eqs-mu}
as an \emph{integration constant}, i.e., as a parameter of the solution
rather than a parameter of the Lagrangian.
The idea that the constant derivative of a field may be important
for the cosmological constant problem has been suggested earlier by
Dolgov~\cite{Dolgov1985,Dolgov1997} and
Polyakov~\cite{Polyakov1991,PolyakovPrivateComm}, where the latter explored
the analogy with the Larkin--Pikin effect~\cite{LarkinPikin1969}
in solid-state physics.

However, instead of the fine-tuning problem of the cosmological constant,
we now have the fine-tuning problem of the integration constant, namely,
the chemical potential $\mu=\mu_0$ that fixes the value $q=q_0$
of the Minkowski equilibrium vacuum
(or \emph{vice versa}, $q_0$ fixing $\mu_0$).
Any other choice of the integration constant ($\mu \ne \mu_0$) leads to a
de-Sitter-type solution~\cite{KlinkhamerVolovik2008b}.
Still, it is possible to show that these de-Sitter-type solutions become
dynamically unstable in a \emph{generalization} of $q$--theory and that
the Minkowski equilibrium vacuum becomes an attractor.

For that purpose, we start from the realization of $q$--theory
in terms of a vector field $A_\beta(x)$ as discussed by Dolgov~\cite{Dolgov1997}
or, equivalently, in terms of an aether-type velocity field $u_\beta(x)$
as discussed by Jacobson~\cite{Jacobson2008}.
The constant vacuum field $q$ then appears~\cite{KlinkhamerVolovik2008a}
as the derivative
of a vector field in the specific solution $u_\beta= \overline{u}_\beta$
corresponding to the equilibrium vacuum,
$q\,g_{\alpha\beta} \equiv \nabla_\alpha\, \overline{u}_{\beta}
\equiv \overline{u}_{\alpha\beta}$.
In this realization, the effective chemical potential
$\mu \equiv d\epsilon(q)/d q$ plays a role only for the equilibrium states
(i.e., for their thermodynamical properties),
but $\mu$ does not appear as an integration constant for the dynamics.
Hence, the fine-tuning problem of the integration constant is overcome,
simply because there is no integration constant.

\begin{figure}[t]
\vspace*{6mm}
\begin{center}
\includegraphics[width=0.35\textwidth]{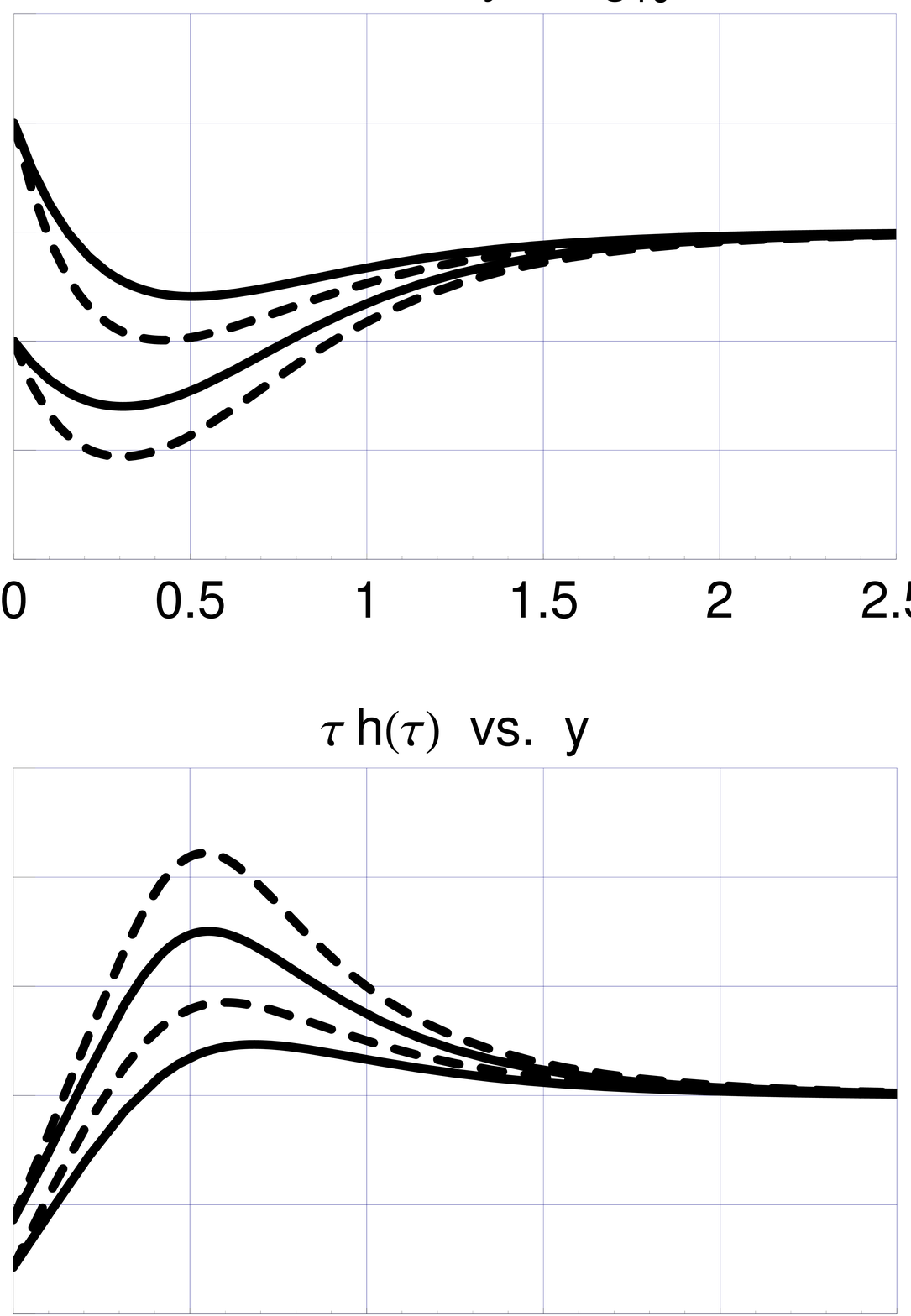}
\end{center}
\vspace*{4mm}
\caption{\textbf{Fig.~2~} Aether-field evolution and Minkowski attractor
in a spatially flat Friedmann--Robertson--Walker universe with $\lambda \equiv
\Lambda_\text{bare}/(E_\text{Planck})^4 = 2$. Top panel: dimensionless
aether-type field component $v \equiv u_0/E_\text{Planck}$ [multiplied by a
factor $\tau^{-1}$] plotted against the logarithm of the dimensionless cosmic
time $y \equiv \log_{10}\,\tau  \equiv \log_{10}\,(t\, E_\text{Planck})$.
Bottom panel: dimensionless Hubble parameter $h \equiv H/E_\text{Planck}$
[multiplied by a factor $\tau$] plotted against $y$. The field equations are
given by Eqs. (5) and (8) in Ref.~\cite{Dolgov1997} for
$\rho_\text{vac}=\Lambda_\text{bare}$ and $\eta_0=-1$: $\ddot{v}+ 3\, h\,
\dot{v} -3\, h^2\, v =0$ and $6\, h^2 =2\,\lambda - (\dot{v})^2 -
3\,(h\,v)^2$, with the overdot standing for differentiation with respect to
$\tau$. The four numerical solutions shown have boundary conditions
$v(1)=1 \pm 0.25$ and $\dot{v}(1)= \pm 0.25$, with the dashed curves referring to
negative $\dot{v}(1)$ [note that the top panel plots $v(\tau)/\tau$, not
$v(\tau)$]. All four numerical solutions approach the Minkowski-spacetime
solution \eqref{eq:asymp_solution} with, for the chosen model parameter
$\Lambda_\text{bare}/(E_\text{Planck})^4 = 2$, a value
$q_0/(E_\text{Planck})^2=1$ appearing dynamically [see top panel].
In fact, the same $t\to\infty$
asymptote is found for all boundary conditions $|v(1)-1| \leq 0.25$  and
$|\dot{v}(1)| \leq 0.25$, which shows that \eqref{eq:asymp_solution} is a
positive attractor. For the theory considered [defined by Eq.~(2) of
Ref.~\cite{Dolgov1997} for $\eta_0=-1$], the Minkowski vacuum is an attractor
because the vacuum compressibility \eqref{eq:chi_0} is positive,
$\chi(q_0)>0$.}
\label{fig:attractor}
\end{figure}

The instability of the de-Sitter solution has been demonstrated by
Dolgov~\cite{Dolgov1997} for the simplest
quadratic choices of the Lagrange density of $u_\beta(x)$
and for an energy scale $E_\text{UV}=E_\text{Planck}$
entering $\Lambda_\text{bare}$ of \eqref{eq:general-epsilon}.
(At this moment, we do not consider the possibility
of having a variable gravitational coupling parameter, so that
we set $K[u_{\alpha\beta}]=K_0=\text{const}$.)
For a spatially flat Robertson--Walker metric with
cosmic time $t$ and scale factor $a(t)$,
the initial de-Sitter-type universe evolves
towards Minkowski spacetime by the following
$t\rightarrow \infty$ asymptotic solution for the aether-type field
$u_\beta=(u_0,u_b)$ and the Hubble parameter $H\equiv [da/dt]/a$:
\beq
u_0(t)\rightarrow q_0\,t \,,\quad
u_b(t)=0 \,,\quad
H(t) \rightarrow 1/t\,,
\label{eq:asymp_solution}
\eeq
where $u_0(t)$ increases linearly with $t$ for constant $q_0$
(the norm of the vector field $u_\beta$ is taken~\cite{KlinkhamerVolovik2008a}
to be unconstrained, different from Ref.~\cite{Jacobson2008}).
Figure~\ref{fig:attractor} shows explicitly the attractor behavior,
with the numerical value of $q_0$ in \eqref{eq:asymp_solution} appearing
\emph{dynamically}.

The following three remarks may help to better understand
the role of \eqref{eq:asymp_solution}.
First, observe that, for finite values of $t$,
the aether-type field approaching \eqref{eq:asymp_solution} does not
correspond to the $q$--theory \emph{Ansatz},
$u_{\alpha\beta} \equiv \nabla_\alpha\, u_{\beta} \ne q\,g_{\alpha\beta}$
for $t < \infty$.
Second, the fact that $H(t)$ in \eqref{eq:asymp_solution}
and Figure~\ref{fig:attractor} drops to zero means that the Robertson--Walker
metric approaches the one of Minkowski spacetime, whereas,
for a positive asymptotic value of $H(t)$,
the metric would have approached the one of de-Sitter spacetime.
Third, the simple model with solution \eqref{eq:asymp_solution}
does not appear to give a realistic description of the present
Universe~\cite{RubakovTinyakov1999} and requires an appropriate modification
(possibly implementing chameleon-type effects~\cite{KhouryWeltman2004}),
but, for the present discussion, the simple model suffices.

It is straightforward to show that the asymptotic solution \eqref{eq:asymp_solution}
also holds for the generalized Lagrangian with a generic
function $\epsilon(u_{\alpha\beta})$ replacing the quadratic term,
as discussed in Sec.~V~D of Ref.~\cite{KlinkhamerVolovik2008a}.
At large cosmic times $t$, the curvature terms
decay as $|R|\sim H^2 \sim 1/t^2$ and the Einstein equations
lead to the nullification of the energy-momentum tensor of the $u_{\beta}$ field:
$T_{\alpha\beta}[u]=0$.
Since the exact expressions on the right-hand-sides of \eqref{eq:asymp_solution}
with $d u_0/dt = H\,u_0$  satisfy the  $q$--theory \emph{Ansatz}
$u_{\alpha\beta} = q\,g_{\alpha\beta}$,
the  energy-momentum tensor is completely expressed by the single constant $q$:
$T_{\alpha\beta}(q) =[\epsilon(q)  -  q\, d\epsilon(q)/d q]\,g_{\alpha\beta}$.
As a result, the equation  $T_{\alpha\beta}(q)=0$
leads to the equilibrium condition \eqref{eq:equil-eqs-q0}
for the Minkowski vacuum and to the equilibrium value $q=q_0$
in \eqref{eq:asymp_solution}. This demonstrates that the compensation of a
large initial vacuum energy density can occur dynamically
and that Minkowski spacetime can emerge
spontaneously, without setting a chemical potential.
In other words, an ``existence proof'' has been given for the conjecture
that the appropriate Minkowski value $q_0$ can result from an attractor-type
solution of the field equations.
The only condition for the Minkowski vacuum to be an attractor
is a positive vacuum compressibility \eqref{eq:chi_0}.
This existence proof is the second of the two most important new results
of the present Letter.

In the previous discussion, we illustrated the compensation of the
``bare'' vacuum energy density by
use of the simplest realizations of the constant vacuum field $q$,
where $q$ follows from derivatives of either the fundamental
gauge field $A_{\beta\gamma\delta}(x)$ or
the fundamental vector field $u_\beta(x)$.
The constant vacuum field $q$ from the four-form field strength
tensor \eqref{eq:Fdefinition} has been discussed earlier in, e.g.,
Refs.~\cite{HenneauxTeitelboim1984,Duff1989}.
But these references consider a quadratic function $\epsilon(F)$,
which can only compensate a $\Lambda_\text{bare}$ value of a particular sign.
Our approach is generic and does not depend on
the particular realization of the ``quinta essentia'' ---
the field $q$ which describes the deep (ultraviolet) quantum
vacuum~\cite{endnote:quinta-essentia}.
The only requirement for $q$ is that it must be a Lorentz-invariant
conserved (i.e., spacetime-independent) quantity in flat Minkowski spacetime.
In addition, an almost arbitrary function $\epsilon(q)$
%allows us to cancel $\Lambda_\text{bare}$ values of both signs;
allows for the cancellation of $\Lambda_\text{bare}$ values of both signs;
see, in particular, the example \eqref{eq:epsilon-example} discussed above.

Finally, it may be of interest to compare
our possible solution of the cosmological constant problem
with that of the unimodular theory of gravity
(see, e.g., Refs.~\cite{Weinberg1988,vanderBij-etal1981,Smolin2009}
and further references therein). From the unimodular theory of gravity,
the cosmological constant of standard general relativity
is obtained as an integration constant
and the Minkowski solution also follows without fine-tuning of the parameters
of the action. As a purely classical theory, unimodular gravity is equivalent to
general relativity, but its extension to the quantum world
can be expected to be different from that of general relativity,
which is at the core of our approach
[the $q$ dependence of the gravitational coupling $K$ in action
\eqref{eq:actionF} is not essential to obtain \eqref{eq:equil-eqs-q0}
and the particular aether-type solution discussed in the present section
already has constant $K$].
Furthermore, the unimodular gravity theory would not allow for
a spacetime-dependent ``cosmological constant'' and, \emph{a forteriori},
would not give an attractor-type solution approaching Minkowski spacetime.

\section{Summary and outlook}\label{sec:Summary}

In this Letter, we have shown that it is possible to find an extension
of the current theory of elementary particle physics (the standard model),
which allows for a Minkowski-spacetime solution with constant fields,
without fine-tuning the extended theory in any way or shape.
For this suggested solution, the cosmological constant $\Lambda_\text{bare}$
from \eqref{eq:general-epsilon}, which includes the
zero-point energy $\Lambda_\text{SM} \sim \pm\, (E_\text{ew})^4$
of the standard model fields,
is completely compensated by the $q$--field that describes
the degrees of freedom of the deep quantum vacuum with energy scale
$E_\text{UV} \gg E_\text{ew}$.

This solves the main cosmological constant problem~\cite{endnote:mainCCP}
and even addresses the next question (also raised in Ref.~\cite{Weinberg1988}):
why is our present Universe close to the Minkowski vacuum or,
in other words, why does Nature prefer flat spacetime?
The answer to this question appears to be:
because the Minkowski equilibrium state is an attractor and the Universe is
moving towards it. We are close to this attractor, simply because our Universe is old.

There remain, however, other problems. Observational cosmology
(see, e.g., Refs.~\cite{Riess-etal1998,Perlmutter-etal1998,Komatsu2008}
and further references therein)
suggests a tiny remnant vacuum energy density $ \rho_{V}$
of the order of $10^{-11}\:\text{eV}^4$.
This, then, leads to the additional cosmic coincidence problem: why is the nonzero
vacuum energy density of the same order as the present matter energy density?
One possible solution~\cite{KlinkhamerVolovik2009b}
of the cosmic coincidence problem may be related to quantum-dissipative effects
during the cosmological evolution of the microscopic field $q(x)$.
In any case, $q$--theory transforms the standard cosmological constant problem
into the search for the proper decay mechanism of the vacuum energy density.

\section*{\hspace*{-4.5mm}ACKNOWLEDGMENTS}
\noindent It is a pleasure to thank A. Dolgov, L. Smolin, and M. Veltman
for helpful comments on an earlier version of this Letter.
GEV is supported in part by the Academy of Finland,
Centers of Excellence Program 2006--2011 and the Khalatnikov--Starobinsky leading scientific school (Grant No. 4899.2008.2).


\begin{thebibliography}{99}
\bibitem{endnote:mainCCP}
As to the terminology of ``main'' cosmological constant problem,
we refer to the required cancellation of a huge initial contribution
to the gravitating vacuum energy density which is
of order $(E_\text{ew})^{4}$ or more;
see also the discussion under Eq.~\eqref{eq:general-epsilon}.
\end{thebibliography}
\end{document}